\documentclass[11pt]{article}
\usepackage[T1]{fontenc}
\usepackage{lmodern}
\usepackage[utf8]{inputenc}
\usepackage[margin=1in]{geometry}
\usepackage{amsmath,amssymb}
\usepackage{graphicx}
\usepackage{array}
\usepackage{booktabs}
\newcolumntype{L}[1]{>{\raggedright\arraybackslash}p{#1}}
\newcommand{\grouphead}[2]{\addlinespace[2pt]\multicolumn{#1}{@{}l}{\emph{#2}}\\[1pt]}
\usepackage{caption}
\usepackage{microtype}
\usepackage[hidelinks]{hyperref}
\hypersetup{
  pdftitle={Passive wearable physiology tracks a state-level material-hardship gradient in resting heart rate},
  pdfauthor={Maria Levchenko; Valeriia Avvakumova; Jane Smorodnikova},
  pdfsubject={Ecological association between state-level material hardship and consumer-wearable resting heart rate},
  pdfkeywords={resting heart rate; material hardship; social determinants of health; consumer wearables; photoplethysmography; digital epidemiology; ecological study}}

\graphicspath{{figures/}}
\newcommand{\secref}[1]{Sec~#1}

\title{\textbf{Passive wearable physiology tracks a state-level material-hardship gradient in resting heart rate}}
\author{Maria Levchenko\thanks{ORCID 0009-0008-2649-0729}, Valeriia Avvakumova, Jane Smorodnikova\\[2pt]\normalsize Welltory Inc.}
\date{}

\begin{document}
\maketitle

\begin{abstract}
\noindent Resting heart rate is an established marker of cardiovascular risk, but population-scale measurement has depended on clinical or survey instruments. We ask whether passively sensed consumer-wearable physiology recovers the socioeconomic gradient established in clinical cohorts. Using 19.1 million quality-filtered photoplethysmography readings from 18,734 opt-in users of the Welltory app, we computed cohort-adjusted mean daytime resting heart rate per US state and related it to a four-component state-level material-hardship composite (uninsurance, food insecurity, utility shutoff, housing insecurity; 41 states with module coverage, 12,497 contributing users). Adjusting for six state health and behaviour indicators, latitude, median age, and density, state resting heart rate tracked hardship at partial Spearman $\rho = +0.74$ (bootstrap 95\% CI $[+0.31, +0.87]$; $[+0.06, +0.79]$ under a conservative two-stage bootstrap that also resamples users within states). The gradient stayed positive under every check: leave-one-state-out ($+0.67$ to $+0.79$), demographic re-weighting, composition and census-region fixed effects ($+0.75$, $+0.78$), split-sample resampling (held-out median $+0.51$ to $+0.56$), and reduced variants such as a heavy-wear subsample ($+0.48$); residuals carry no detectable spatial autocorrelation. It was absent in two other metrics from the same panel and not matched by income or inequality under the same adjustment. Users in the five highest-hardship states averaged $+1.33$ bpm over those in the five lowest. Passive consumer-wearable physiology tracks, at the area level, a gradient previously established primarily in clinical and cohort studies, without surveying the users themselves. The composite was selected in exploratory analysis, after a pre-specified broader precarity index gave a weaker association ($+0.29$); the association is cross-sectional and ecological.
\end{abstract}

\section*{1\quad Introduction}

Resting heart rate (RHR) is among the most reproducible and prognostically informative physiological signals in cardiovascular epidemiology. Across large cohorts, higher resting heart rate is associated in a continuous, graded fashion with cardiovascular and all-cause mortality \cite{Aune2017,Cooney2010,Woodward2014}. It is also an accessible index of sympathetic (chronotropic) tone and chronic physiological load, rising with sustained sympathetic activation. Yet population-scale measurement of RHR has historically depended on clinical examination or survey-linked instruments, which are episodic, costly, and lag the conditions they aim to capture. The result is a persistent gap between how precisely RHR can be measured in an individual and how coarsely it can be observed across a population.

That population-scale gap matters because RHR carries a social signature. In the RECORD cohort, resting heart rate rose monotonically with socioeconomic disadvantage --- by approximately $+0.9$, $+1.8$, and $+3.6$ bpm across three increasing categories of combined individual and neighbourhood disadvantage, relative to the least-disadvantaged category \cite{Chaix2011}. The individual-level association between socioeconomic position and resting heart rate is, in other words, established epidemiology. Its proposed mechanism is equally well developed: lower socioeconomic status is accompanied by chronic exposure to stressors, sustained autonomic activation, and delayed cardiovascular recovery, a pattern interpreted through the framework of allostatic load --- the cumulative physiological cost of repeated or prolonged stress responses \cite{McEwen1998,Steptoe2002}.

Whether this social signature is legible in \emph{passively sensed} physiology, at population scale, remains open. Two adjacent literatures bracket the question --- area-level studies link socioeconomic deprivation to cardiovascular \emph{disease} endpoints, and consumer-wearable studies aggregate resting heart rate to large geographies and validate it against external \emph{surveillance} --- and a single prior study has related area-level deprivation to heart rate directly, in the opposite direction to ours. None has related a passively measured, area-aggregated resting heart rate to a direct measure of material hardship. That intersection is the gap this paper fills.

\subsection*{1.1\quad Related work}

\textbf{Area-level deprivation and cardiovascular endpoints.} Small-area socioeconomic deprivation is an established predictor of cardiovascular outcomes. In a linkage study of nearly two million people, deprivation predicted the incidence of a wide range of cardiovascular diseases \cite{Pujades2014}, and it predicts cardiovascular mortality in older cohorts \cite{Ramsay2015}; deprivation has likewise been associated with cumulative physiological burden, such as higher allostatic load in more deprived areas \cite{Ribeiro2019}. These studies establish that area-level disadvantage leaves a cardiovascular imprint, but their endpoints are disease events, mortality, or multi-marker load indices --- not resting heart rate itself.

\textbf{Population-scale wearable resting heart rate.} Consumer wearables have made resting heart rate observable at population scale. Aggregated wearable RHR has been shown to track influenza-like-illness activity at the US-state level, establishing that passively sensed RHR carries real population signal and can be validated against an external indicator \cite{Radin2020}. This literature validates wearable RHR against infectious-disease surveillance and, in related work, against physiological states in diagnosed individuals --- but not against a direct area-level measure of material hardship.

\textbf{Area-level heart rate and deprivation.} We are aware of one prior study relating area-level deprivation directly to heart rate: the Baltimore Study of Black Aging reported \emph{decreasing} heart rate with greater neighbourhood disadvantage \cite{Allan2024}. It is a longitudinal analysis of change across two waves roughly three years apart in a single-city cohort of 317 older Black adults; its design, population, and direction differ from the present cross-sectional, general-panel estimate, and we reconcile the two in the Discussion (\secref{5.3}).

The intersection of these threads --- a passively collected, consumer-wearable resting heart rate, aggregated to the area level, related cross-sectionally to a direct composite of material hardship --- is, to our knowledge, not previously reported, and it is the gap we address.

Using 19.1 million quality-filtered photoplethysmography readings from 18,734 opt-in users of Welltory, a consumer health application, across all 50 US states and the District of Columbia, we compute a cohort-adjusted state-level mean resting heart rate and relate it to a state-level composite of four federal material-hardship indicators --- health-insurance non-coverage, food insecurity, utility-shutoff rate, and housing insecurity. After adjustment for state-level health, behavioural, demographic, and geographic covariates, state mean resting heart rate tracks material hardship, across the 41 states with complete social-determinant coverage (12,497 contributing users), at a partial Spearman correlation of $\rho = +0.74$. To our knowledge, this is the first demonstration of a positive, cross-sectional, \emph{area-level} association between \emph{resting heart rate} and \emph{material hardship}, and the first to establish it from \emph{passive consumer-wearable} data. The contribution is methodological as much as substantive: it shows that passively sensed physiology tracks, at population scale, a social gradient in resting heart rate previously established primarily in clinical and cohort studies --- without administering any survey to the users themselves (the hardship indicators are themselves federal survey products). We frame the result for what it is --- an ecological, cross-sectional association among an opt-in panel, consistent with established socioeconomic cardiovascular epidemiology, and not a causal or individual-level claim.

\section*{2\quad Materials and methods}

\subsection*{2.1\quad Data source and panel}

Physiological data derive from the passive-physiology panel of \textbf{Welltory}, a consumer health application whose users record heart-rate measurements through photoplethysmography (PPG), captured either from the smartphone camera or from the Apple Watch optical sensor. Users were assigned to a US state from mobile network/device location data associated with their readings, not from self-reported residence; the source panel covers 2024--2025. After quality control, the working panel comprised \textbf{18,734 users} contributing \textbf{19.1 million} quality-filtered waking readings across all 50 states and the District of Columbia. Each user was assigned to exactly one state; users whose readings could not be resolved to a single state were excluded, so the panel contains no multi-state records.

The panel is opt-in and not population-representative: users self-select into the application, and their demographic and geographic distribution does not match the US adult population (analysed in the Limitations and in \secref{4}). All analyses and claims are therefore scoped to patterns among these users and their consistency with external indicators, not to the general population.

\subsection*{2.2\quad Resting heart rate (``BPM day-only'')}

Each physiological reading is a short (60--120 s) photoplethysmography recording collected through the Welltory application. Apple Watch recordings, which supply $\approx$98\% of the panel, are scheduled opportunistically through the day by the application rather than initiated by the user; smartphone-camera recordings, the small remainder, are by their nature user-initiated, since they require the user to hold a finger to the lens. In this US 2025 panel these recordings are predominantly from the Apple Watch optical sensor ($\approx$98\% of recordings; the median contributing user's readings are essentially all Apple Watch, and only about 2\% of users record solely by smartphone camera). Unlike a passively and continuously sampled heart rate, each recording captures pulse-to-pulse (inter-beat) intervals --- the photoplethysmographic analogue of the R--R interval, which is an electrocardiographic feature --- and therefore yields both a heart rate and heart-rate-variability metrics; a user contributes a small number of such readings per waking day (median 2 day-only readings per active day), not a single device-computed daily value. Each recording yields one heart-rate value, stored in the panel as the per-reading field \texttt{bpm\_1\_0}. Our physiological measure, denoted \textbf{BPM day-only}, is the per-user median of \texttt{bpm\_1\_0} over that user's daytime recordings (those tagged \texttt{time\_of\_day\_1\_0 == "day"}), requiring at least 20 such readings for a user to contribute. Because each recording is brief and opportunistic, no single reading is a resting-state measurement, and any given one may coincide with activity; we therefore do not treat individual readings as resting values. Instead, the per-user daytime \emph{median} over many such readings --- 10.2 million day-only readings in 2025 across the qualifying users, a median of 373 per user (interquartile range 94--897) spread over a median of 168 days --- and robust to the minority taken during activity, serves as our internal estimate of a typical \textbf{waking resting heart rate}: we expect most opportunistic daytime readings to be taken at rest or during sedentary activity --- an assumption we cannot verify without activity-state data (\secref{6}) --- so their per-user central tendency plausibly approximates a resting level. We use the term ``resting heart rate'' throughout to denote this internally derived estimate. The \texttt{bpm\_1\_0} value is derived from each raw recording by Welltory's own signal-processing pipeline, and is distinct both from a clinician-measured seated resting heart rate and from the nocturnal-minimum ``resting heart rate'' reported by some wearables; we have not validated it against a conventional resting-heart-rate instrument, and we state the construct correspondence as an explicit assumption in the Limitations (\secref{6}).

To remove composition effects arising from the panel's age and sex structure, each user's value was expressed as a deviation from the mean of their sex $\times$ age-bucket cohort (``cohort adjustment''), and these deviations were then averaged to the state level. Age and sex are user-reported in the application profile rather than verified, so cohort assignment inherits any misreporting; users missing either field cannot be assigned a cohort and do not contribute to a state mean. A state was included only if at least 75 users contributed to its mean, ensuring each state estimate rests on a non-trivial sample.

The resting-heart-rate estimate is computed over the \textbf{2025 measurement window}. The panel spans 2024--2025, but 2024 data are used only to assess year-over-year rank stability (\secref{4.4}); the cross-sectional association reported here is a 2025 estimate. Where the panel period ``2024--2025'' appears, it denotes coverage of the source panel, not the window of the reported estimate.

Alongside resting heart rate we derive two ancillary physiological metrics from the same panel, used only as metric-specificity and device-mix comparisons (\secref{3.4}, \secref{4.3}), not for any headline claim: \textbf{Intensity}, the share of a user's readings classified as a high-arousal state, and \textbf{Drain}, a day-over-day energy-depletion (recovery) metric. Both are cohort-adjusted and averaged to the state in the same way as BPM day-only.

\subsection*{2.3\quad Contributing samples at each analysis tier}

The number of users contributing to a state mean depends on the measurement-availability filter, and we report the relevant sample at each tier rather than a single global $N$. For the headline four-component analysis (Composite4, \secref{2.4}), which is defined over the 41 states with full social-determinant module coverage, \textbf{12,497 users} (each with $\geq$20 readings) across the 41 states that each had $\geq$75 contributing users made up the state means. For the two-component variant (Composite2), defined over all 51 jurisdictions, \textbf{17,715 users} contributed. The unit of statistical analysis is the state ($n = 41$ for Composite4; $n = 51$ for Composite2); the user counts are the aggregation inputs to each state mean, not the sample over which the correlation is computed.

The reliability analyses (\secref{4.4}) use their own cohorts, which are not the samples on which the headline association is estimated: a 17,196-user and a 15,476-user cohort under a stricter sustained-measurement rule ($\geq$50 days) for the state-ranking and wearing-intensity checks, and, for the user-level split-half, the 17,379 users with at least 10 day-only readings in each half of the split.

\subsection*{2.4\quad Material-hardship composite}

Material hardship was operationalised as a z-scored composite of four state-level social-determinant indicators drawn from US federal sources: health-insurance non-coverage (US Census / American Community Survey), food insecurity (US Department of Agriculture, Economic Research Service), and utility-shutoff rate and housing-insecurity rate (US Centers for Disease Control and Prevention, PLACES social-determinants-of-health module). All variables --- outcome, composite components, and controls --- with their roles and federal sources are summarised in Table~\ref{table1}. The four-component composite (``\textbf{Composite4}'') is defined for the 41 jurisdictions (40 states and the District of Columbia) with PLACES SDOH-module coverage; for brevity we refer to these as ``41 states'' elsewhere. A two-component variant (``\textbf{Composite2}'': uninsurance and food insecurity), available for all 51 jurisdictions, is reported as a national-coverage supplement.

The ten jurisdictions whose health departments did not field the 2022 module are Colorado, Florida, Oregon, Pennsylvania, South Dakota, Tennessee, Texas, Vermont, Washington, and Wyoming. Their absence is not neutral with respect to either variable: on the two components available nationally they are slightly \emph{higher} hardship than the 41 (mean Composite2 $+0.09$ vs $-0.02$; mean uninsurance 8.5\% vs 7.3\%, and they include Texas, the highest-uninsurance state at 16.7\%), while their cohort-adjusted resting heart rate is \emph{lower} (mean $-0.18$ vs $+0.12$ bpm). They therefore sit against the gradient, and the 41-state subset is the more favourable of the two samples --- visible directly in Composite2, which gives $+0.63$ on the 41 states and $+0.57$ on all 51 (S1 Table). We report both throughout rather than only the covered subset.

Components were standardised to z-scores before averaging. Because module coverage differs across components, the two PLACES-derived components (utility shutoff, housing insecurity) were standardised to the moments of the 41 states for which they are defined, while the two broader-coverage components (uninsurance, food insecurity) were standardised to the 51-state moments. A composite formed as the mean of z-scores is therefore not strictly rank-invariant to the standardisation base; we adopted this base deliberately for consistency of each component with its own coverage, and confirmed the choice is immaterial --- re-standardising all four components to the 41-state base leaves the partial correlation essentially unchanged ($\rho = +0.746$ vs $+0.740$; the two composites rank-correlate at 0.997).

The hardship indicators and covariates are drawn from recent reference periods (2022--2024, depending on the series --- Table~\ref{table1}), while resting heart rate is estimated over 2025; because state-level hardship rankings are generally stable year to year, this offset is unlikely to materially affect a rank-based cross-sectional analysis (though for the two single-year 2022 components we cannot verify this directly, \secref{6}).

\begin{table}[ht]
\centering\footnotesize
\caption{Variables and sources. Composite4 $=$ z-scored mean of the four components (41 states with SDOH coverage); Composite2 $=$ z-scored mean of \% uninsured $+$ food insecurity (51 jurisdictions).}
\label{table1}
\begin{tabular}{L{3.0cm} L{6.4cm} L{5.6cm}}
\toprule
\textbf{Variable} & \textbf{Definition} & \textbf{Source (agency $\cdot$ series $\cdot$ year)} \\
\midrule
\grouphead{3}{Outcome}
BPM day-only & Per-user median daytime resting HR ($\geq$20 readings), cohort-adjusted (sex$\times$age), averaged to state & Welltory passive-PPG panel $\cdot$ 2025 \\
\grouphead{3}{Ancillary metrics --- specificity checks only}
Intensity & Share of readings in a high-arousal state & Welltory $\cdot$ 2025 \\
Drain & Day-over-day energy-depletion (recovery) metric & Welltory $\cdot$ 2025 \\
\grouphead{3}{Hardship composite components}
\% uninsured & Population without health insurance & US Census / ACS $\cdot$ S2701 $\cdot$ 2024 \\
Food insecurity & State food-insecurity rate & USDA ERS $\cdot$ CPS-FSS $\cdot$ 3-yr avg 2022--2024 \\
Utility shutoff & Utility-shutoff rate & CDC PLACES $\cdot$ BRFSS SDOH module $\cdot$ 2025 release \\
Housing insecurity & Housing-insecurity rate & CDC PLACES $\cdot$ BRFSS SDOH module $\cdot$ 2025 release \\
\grouphead{3}{Controls (9)}
Health and behaviour (6) & Obesity, diagnosed diabetes, hypertension, physical inactivity, current smoking, self-rated fair/poor health & CDC PLACES $\cdot$ BRFSS $\cdot$ 2025 release \\
Latitude & State centroid latitude & Geographic \\
Median age & State median age & US Census / ACS $\cdot$ 2024 \\
Population density & People per square mile & US Census $\cdot$ 2024 \\
\grouphead{3}{Context --- not in the composite or the control set}
Gini; unemployment; bankruptcy & Income inequality; 2025 unemployment; FY2025 bankruptcy filings per 100k & ACS B19083 $\cdot$ BLS LAUS $\cdot$ US Courts \\
\bottomrule
\end{tabular}

\vspace{2pt}
\begin{minipage}{\textwidth}\raggedright
{\footnotesize \emph{Reference periods.} CDC PLACES measures are small-area model-based estimates from the 2025 PLACES release: the six health/behaviour controls from 2023 BRFSS, and utility shutoff and housing insecurity from the 2022 BRFSS Social Determinants and Health Equity optional module --- administered by only a subset of jurisdictions, which is why Composite4 is limited to 41 jurisdictions (40 states and DC). USDA food insecurity is a three-year pooled prevalence (2022--2024). ACS figures are 2024 one-year estimates; BLS unemployment is 2025; US Courts bankruptcy is FY2025.}
\end{minipage}
\end{table}

\subsection*{2.5\quad Covariates}

To distinguish a hardship--RHR association from confounding by population health, behaviour, geography, and demography, we adjusted for nine state-level covariates: six CDC PLACES health and behaviour prevalences --- obesity, diagnosed diabetes, hypertension, physical inactivity, current smoking, and self-rated fair-or-poor health --- together with latitude, median age, and population density. These six health and behaviour measures include both plausible confounders and plausible mediators of a hardship--RHR relationship; we report the association under both readings (\secref{3}, \secref{4}) and show that the conclusion does not depend on their classification.

\subsection*{2.6\quad Statistical analysis}

The primary estimate is the partial Spearman rank correlation between state mean BPM day-only and the hardship composite, adjusting for the nine covariates above. It is computed in rank space: outcome, exposure, and each covariate are rank-transformed, the ranked outcome and exposure are each regressed on the ranked covariates by ordinary least squares, and the partial correlation is the Pearson correlation of the two residual vectors. We report the rank rather than the linear (Pearson) correlation because it is robust to non-linearity and to the influence of individual states, consistent with the decision not to make ordinal state-ranking claims (\secref{4.4}). We report the marginal (unadjusted) alongside the partial (adjusted) correlation; in our data the partial exceeds the marginal.

Uncertainty is quantified two ways. The nonparametric bootstrap resamples the 41 states with replacement (2{,}000 iterations, seed 0), recomputing the full partial correlation --- covariate regression included --- within each resample, and takes the 2.5th and 97.5th percentiles. The Bayesian interval places a flat prior on the Fisher $z$ transform of the partial correlation, the standard non-informative choice: given $\hat{\rho}$ on $n$ states with $k$ covariates, $z = \operatorname{atanh}(\hat{\rho}) \sim \mathcal{N}(\operatorname{atanh}(\rho),\, 1/(n-k-3))$, and the posterior on $z$ is transformed back to $\rho$ ($n = 41$, $k = 9$, 20{,}000 posterior draws). The bootstrap is the wider of the two because it propagates the covariate regression and the resampling of states, which the closed-form posterior conditions on. Both, however, treat each state mean as known without error, while those means rest on 75 to 1,668 users apiece. We therefore also ran a \textbf{two-stage bootstrap} that resamples the 41 states and then resamples users with replacement within each drawn state, recomputing the sex $\times$ age-bucket cohort means and re-aggregating the state means inside every iteration, so that within-state measurement error and uncertainty in the cohort adjustment are both propagated. This scheme is conservative by construction: each resampled state mean carries a fresh draw of sampling noise on top of the noise already present in the observed mean, so the resampled correlation is attenuated relative to the point estimate and its interval should be read as a bound rather than a calibrated interval. Analyses were run in Python 3.12 with NumPy, SciPy, and pandas; all seeds are fixed and stated in the released code (\secref{7}).

To probe out-of-sample behaviour we used a pre-specified 31/10 state train--holdout split, with seed and split fraction set before the construct refinement (seed 42), and additionally report a nested cross-validation in which component selection is repeated inside each fold, so that the held-out estimate does not benefit from selection on the full sample (\secref{4.6}). Demographic sensitivity was assessed by re-computing the state means under direct standardisation to external population margins, using three standardisation estimators (user-weighted, ACS-weighted, and cohort-residualised; \secref{3.3}).

The pre-specified analysis used a broader economic-precarity index (EPI) combining unemployment, uninsurance, rent burden, and food insecurity. Following it, we carried out an exploratory construct-refinement focused on \emph{direct} experiences of material hardship, which produced Composite4. Because the construct was chosen after seeing the data, we quantify the resulting optimism directly rather than assert it away: the nested cross-validation repeats component selection inside each fold (\secref{3.2}, \secref{4.6}) and is reported alongside the pre-specified-split result.

\section*{3\quad Results}

The pre-specified EPI analysis produced a positive but modest partial Spearman association with the resting-heart-rate estimate ($\rho = +0.29$, 51 jurisdictions). Exploratory examination indicated that this index mixed \emph{direct} experiences of hardship with broader labour- and housing-market conditions --- in particular, state rent burden also captures high-cost housing markets and did not behave as a unidirectional deprivation indicator. We therefore evaluated a narrower, direct-hardship construct --- Composite4 (uninsurance, food insecurity, utility shutoff, and housing insecurity) --- and report its exploratory association below; the candidate-by-candidate rationale is in the supplement (S1 Table).

\subsection*{3.1\quad State resting heart rate tracks material hardship}

Across the 41 states with full social-determinant coverage, state mean resting heart rate was strongly and positively associated with the four-component hardship composite. Our primary estimand is the full-sample partial Spearman correlation across these 41 states, adjusting for the nine covariates. It was $\rho = +0.74$ (nonparametric bootstrap 95\% confidence interval $[+0.31, +0.87]$; Bayesian 95\% credible interval $[+0.53, +0.87]$, posterior probability that $\rho > 0$ exceeding 0.999). The two intervals answer different questions: the Bayesian interval quantifies uncertainty in the residual correlation conditional on the covariate adjustment, while the bootstrap additionally propagates that adjustment and the resampling of states, which is why it is wider and why we quote it first.

Neither interval, however, accounts for the fact that the state means are themselves estimates. Propagating that error with the two-stage bootstrap (\secref{2.6}) --- resampling states, then users within states, recomputing the cohort adjustment each time --- widens the interval to $[+0.06, +0.79]$, with 98.5\% of iterations positive. It also pulls the median of the resampled distribution down to $+0.52$, which is the attenuation expected when every state mean carries a second, artificial realisation of its own sampling noise; the two-stage interval is therefore a conservative bound rather than a calibrated replacement for $[+0.31, +0.87]$. Classical measurement error runs the other way --- it biases the observed $+0.74$ toward zero, so a disattenuated estimate would be larger. Read together, these bracket the honest position: the sign of the association is robust under every scheme we ran, while its magnitude is less precisely determined than the state-level interval alone implies.

The unadjusted (marginal) correlation was $+0.64$: adjustment for the covariates \emph{increased} rather than attenuated the association. This suppression is a joint (multivariate) effect, not an artefact of any single covariate --- across all 511 non-empty subsets of the nine covariates the partial correlation remains positive, ranging $+0.39$ to $+0.74$ (\secref{4.1}). The full nine-covariate set sits at the top of that range, so the specification we headline is the most favourable of the 511; the relevant fact is that the floor is $+0.39$, not that the ceiling is ours. The six health and behaviour covariates are the dominant net suppressor: controlling them alone raises the correlation from $+0.64$ to $+0.69$. The three demographic and geographic controls (latitude, median age, population density) alone move it slightly the other way, to $+0.59$, and contribute further \emph{joint} suppression only once the health covariates are also present, giving $+0.74$.

\begin{figure}[ht]
\centering
\includegraphics[width=0.82\textwidth]{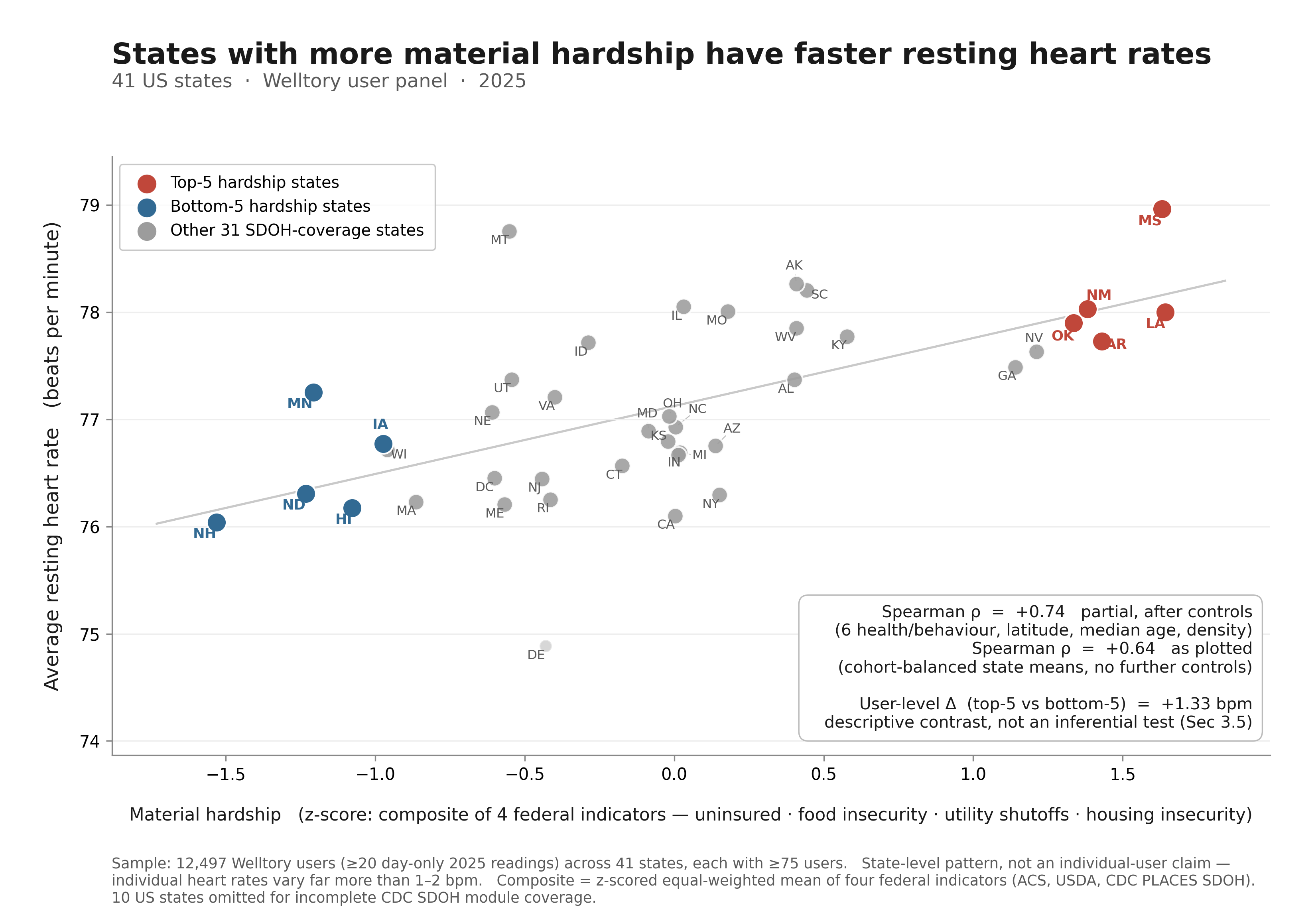}
\caption{State-level material-hardship composite (Composite4) against cohort-adjusted mean resting heart rate (BPM day-only), 41 states with CDC PLACES SDOH module coverage, with fit and confidence band; partial Spearman $\rho = +0.74$ annotated. State abbreviations are shown to aid orientation; individual state positions are sample-limited and we make no ordinal state-ranking claims (\secref{4.4}).}
\label{fig1}
\end{figure}

Read across specifications, the association is robustly positive (Table~\ref{table2}). Comparable full-sample estimates cluster tightly around the headline --- leave-one-state-out $+0.67$ to $+0.79$, demographic re-weighting $+0.67$ to $+0.74$ (\secref{3.3}, \secref{4.1}) --- while deliberately reduced variants run lower, roughly $+0.48$ to $+0.57$ (a single-covariate adjustment and a heavy-wearer subsample). The two-component composite, the one specification carried over all 51 jurisdictions, is $+0.57$ there and $+0.63$ on the 41 module-covered states: the gradient is national, and somewhat stronger in the covered subset than outside it (\secref{2.4}). We headline the four-component composite on construct grounds, as the more complete operationalisation of material hardship, and report the national two-component figure alongside it throughout. Composite2 doubles as a robustness anchor against the two single-year (2022) CDC PLACES components: its indicators --- uninsurance (ACS) and food insecurity (USDA) --- are not drawn from the BRFSS social-determinants module, so its $+0.57$ shows the gradient does not depend on those two components or on their 2022 vintage. In substantive terms, states whose users show higher resting heart rate are also, systematically, states with greater material hardship, after accounting for population health, behaviour, geography, and demography (Fig~\ref{fig1}).

\begin{table}[ht]
\centering\footnotesize
\caption{Results ladder: the association across specifications. All estimates are partial Spearman $\rho$ between state mean BPM day-only and the hardship composite unless noted.}
\label{table2}
\begin{tabular}{L{5.6cm} l L{5.0cm} r}
\toprule
\textbf{Specification} & \textbf{$\rho$} & \textbf{95\% interval / detail} & \textbf{n} \\
\midrule
\grouphead{4}{Full sample, 41 states}
Marginal (unadjusted) & $+0.64$ & --- & 41 \\
\textbf{Partial, 9 controls --- headline} & $\mathbf{+0.74}$ & bootstrap $[+0.31, +0.87]$; Bayesian $[+0.53, +0.87]$; $P(\rho{>}0) > 0.999$ & 41 \\
\quad two-stage bootstrap (states $+$ users) & $+0.74$ & $[+0.06, +0.79]$; 98.5\% of draws $>0$; conservative & 41 \\
$+$ Race/ethnic composition (11 ctrl) & $+0.752$ & bootstrap $[+0.31, +0.88]$ & 41 \\
$+$ Census-region fixed effects (12 ctrl) & $+0.776$ & bootstrap $[+0.29, +0.90]$ & 41 \\
Demographic re-weighting (3 estimators) & $+0.67$ to $+0.74$ & all intervals exclude zero & 41 \\
\grouphead{4}{Influence and alternative samples}
Leave-one-state-out (41 fits) & $+0.67$ to $+0.79$ & --- & 41 \\
Leave-one-region-out (4 fits) & $+0.62$ to $+0.72$ & --- & 27--34 \\
Within-state heavy-wear subset & $+0.475$ & attenuation partly reliability at half the sample & 41 \\
Composite2 (national supplement) & $+0.573$ & bootstrap $[+0.21, +0.74]$ & 51 \\
\grouphead{4}{Held out --- internal resampling, same 41 states}
Pre-specified split (seed 42) & $+0.83$ & predicted-vs-actual held-out ranks & 10 \\
50 random splits, Composite4 fixed & $+0.56$ med. & pred-vs-actual $>0$ in 49/50; marginal $>0$ in 50/50 & $10{\times}50$ \\
Nested CV, selection inside each fold & $+0.51$ med. & positive in 98\% of folds; optimism $-0.055$ & $10{\times}50$ \\
\bottomrule
\end{tabular}
\end{table}

\subsection*{3.2\quad The association is stable across internal split-samples}

The relationship was not an artefact of fitting all 41 states at once. Across 50 random 31/10 train--holdout splits, the held-out predictions tracked the actual held-out ranks at a median predicted-versus-actual correlation of $+0.56$, positive in 49 of 50 splits; the held-out marginal correlation was positive in all 50 (two distinct quantities we keep separate). The pre-specified split (seed 42) gave $+0.83$ --- a favourable draw within this distribution.

\begin{figure}[ht]
\centering
\includegraphics[width=0.9\textwidth]{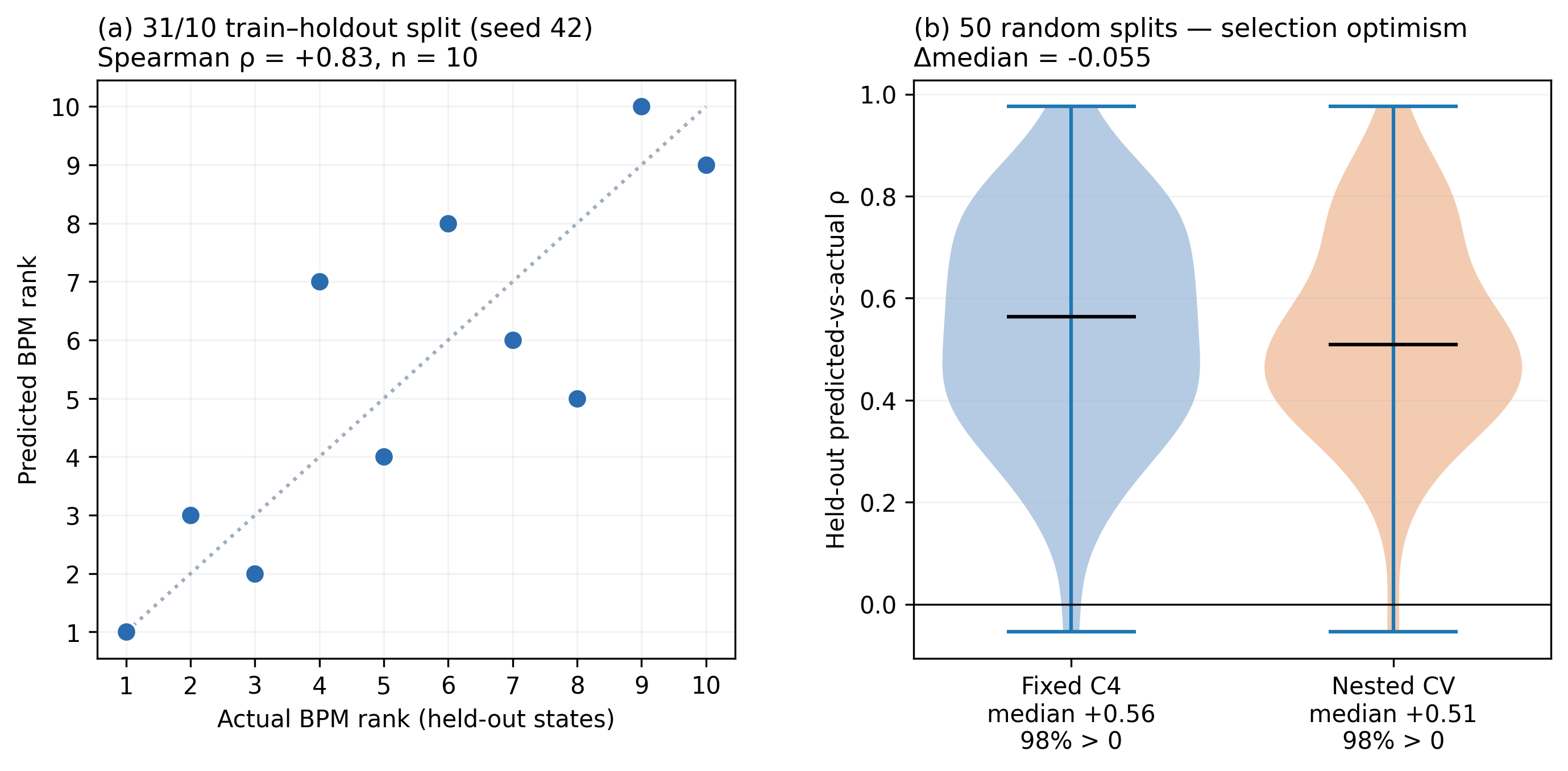}
\caption{Internal split-sample stability (the 50 splits reuse the same 41 states; this is internal resampling, not independent external validation). (a) Predicted-versus-actual held-out state ranks for the pre-specified 31/10 split (seed 42), $\rho = +0.83$. (b) Distribution of held-out predicted-versus-actual $\rho$ over 50 random splits under fixed-Composite4 selection (median $+0.56$) versus nested cross-validation with selection repeated inside each fold (median $+0.51$; the nested estimate is lower by 0.055), both positive in 98\% of folds.}
\label{fig2}
\end{figure}

The chronology bounds what that $+0.83$ means. The split, its seed, and its fraction were fixed before the construct refinement, but the ten held-out states were not withheld from it: Composite4's components were chosen while all 41 states were visible (the candidates in S1 Table are scored on all 41), so $+0.83$ is not a clean out-of-sample test of the construct. Two facts limit the consequence. On this split, redoing the selection using only the 31 training states picks the same composite and reproduces $+0.83$ exactly, so the held-out states did not change which composite won. And repeating component selection inside every fold (nested cross-validation) costs 0.055 of the median held-out $\rho$, lowering it to $+0.51$ while leaving the sign positive in 98\% of folds (Fig~\ref{fig2}). That 0.055 covers the restricted selection step reproduced inside the folds; it does not quantify the optimism from the wider exploratory search that produced Composite4, which remains unknown (\secref{4.6}). This is internal validation, not confirmation.

\subsection*{3.3\quad The association survives demographic re-weighting}

Re-computing the state means under direct standardisation to external population margins left the association essentially unchanged. Three estimators were used: user-weighted standardisation (partial $\rho = +0.667$, bootstrap $[+0.24, +0.83]$), standardisation to ACS state age-by-sex margins ($+0.681$, $[+0.28, +0.86]$), and the cohort-residualised estimator carried through the rest of the paper ($+0.740$). The first two re-weight each state's users toward an external population structure and cost roughly $0.06$ of correlation for doing so; all three exclude zero. The gradient is therefore not an artefact of the panel's age and sex composition, though the modest attenuation under external weighting is worth noting: the more the panel is forced toward the general population's demographic structure, the slightly weaker the estimate.

\subsection*{3.4\quad The gradient is specific to resting heart rate and to hardship}

\begin{figure}[ht]
\centering
\includegraphics[width=0.9\textwidth]{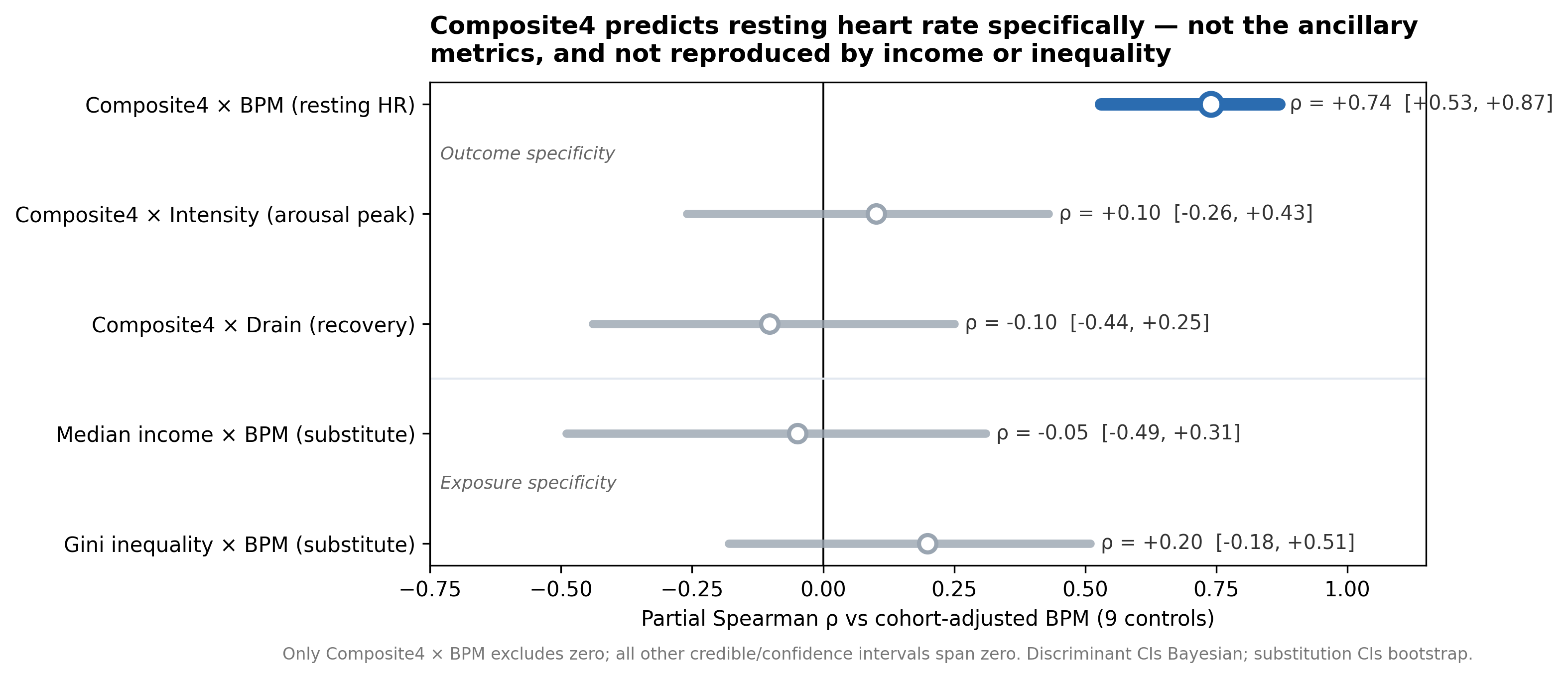}
\caption{Specificity of the gradient. Partial Spearman $\rho$ (with credible intervals and a zero reference line) of the hardship composite against resting heart rate ($+0.74$) versus the two ancillary metrics, Intensity ($+0.10$) and Drain ($-0.10$); and of resting heart rate against median household income ($-0.05$) and income inequality / Gini ($+0.20$). Only the hardship composite $\times$ resting-heart-rate cell departs from zero.}
\label{fig3}
\end{figure}

Two specificity checks distinguish the finding from more generic alternatives. First, \textbf{metric specificity}: the hardship composite tracked resting heart rate (partial $\rho = +0.74$) but neither of the two ancillary metrics derived from the same panel --- a recovery measure (Drain, partial $\rho = -0.10$) and an arousal-peak measure (Intensity, $+0.10$), each with a credible interval spanning zero --- indicating the association is specific to resting heart rate rather than a general property of any signal the panel produces. Drain provides the cleaner discriminant evidence because, like resting heart rate, it passes that check (\secref{4.3}); Intensity's null is consistent but weaker, since Intensity itself fails that check. Second, \textbf{hardship survives the adjustment that removes income's association}. Both are associated with resting heart rate before adjustment, in the expected directions: hardship at a marginal $\rho = +0.64$, median household income at $-0.513$. Under the nine-covariate adjustment the hardship association strengthens to $+0.74$ while income's falls to $-0.05$; substituting income inequality (Gini) gives $+0.20$, with a credible interval spanning zero. The comparison is therefore between two exposures under one adjustment set, not a claim that state income is unrelated to resting heart rate --- it plainly is, marginally. What survives adjustment for population health, behaviour, geography, and demography is the direct-hardship bundle (uninsurance, food insecurity, utility and housing precarity), not average income or its dispersion (Fig~\ref{fig3}). Because six of the nine covariates are themselves socioeconomically patterned, part of income's attenuation is shared variance with them; hardship's survival of the same adjustment is the contrast worth reading.

\subsection*{3.5\quad An individual-level anchor}

\begin{figure}[ht]
\centering
\includegraphics[width=0.82\textwidth]{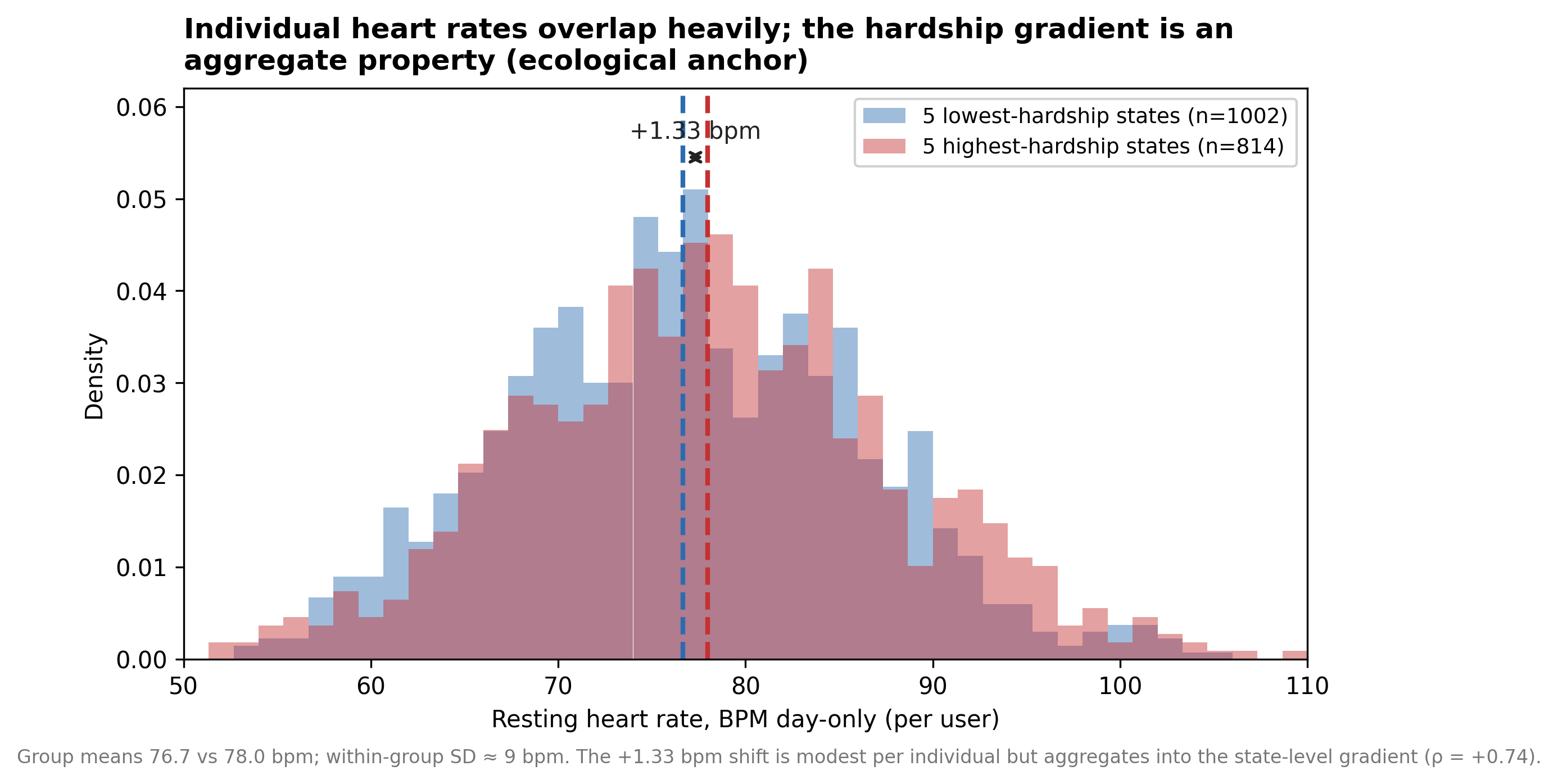}
\caption{Ecological anchor. Overlaid distributions of per-user resting heart rate (raw day-only values, not cohort-adjusted) for users in the five highest- versus five lowest-hardship states. The distributions overlap heavily (within-state between-user SD $\approx$ 9 bpm) around a small mean shift ($+1.33$ bpm) --- a modest individual difference that aggregates into the strong between-state gradient.}
\label{fig4}
\end{figure}

The primary estimate is ecological (state-level; \secref{5}, \secref{6}). To anchor it at the individual level, we compared users directly. Users in the five highest-hardship states had a mean resting heart rate of 77.99 bpm, versus 76.66 bpm for users in the five lowest --- a difference of $+1.33$ bpm ($n = 814$ and $1{,}002$ users). This is a descriptive contrast, not an inferential test: hardship is assigned at the state level and only five states enter each group, so the users are not independent observations and a user-level significance test would understate that clustering. Against a within-state, between-user standard deviation of roughly 9 bpm, $+1.33$ bpm is a modest difference at the level of any individual, and it aggregates into the strong between-state gradient reported above --- the expected relationship between a small individual effect and a large ecological one (\secref{6}; Fig~\ref{fig4}). Note that this user-level contrast is a different quantity from the difference between the corresponding state means ($+1.54$ bpm on raw values); we report the user-level figure here because the anchor is meant to be an individual-level one.

\section*{4\quad Robustness}

We subjected the primary association to influential-observation sensitivity, alternative specifications, spatial dependence and population composition, a device-mix sensitivity check, reliability at two grains, a structurally corrected wearing-intensity check, and the multiple-comparisons context. It is stable across the first six. The reliability and multiple-comparisons analyses set the boundaries of what we claim, and we report them in the same detail as the confirmatory checks.

\subsection*{4.1\quad Influential observations and alternative specifications}

\begin{figure}[ht]
\centering
\includegraphics[width=0.9\textwidth]{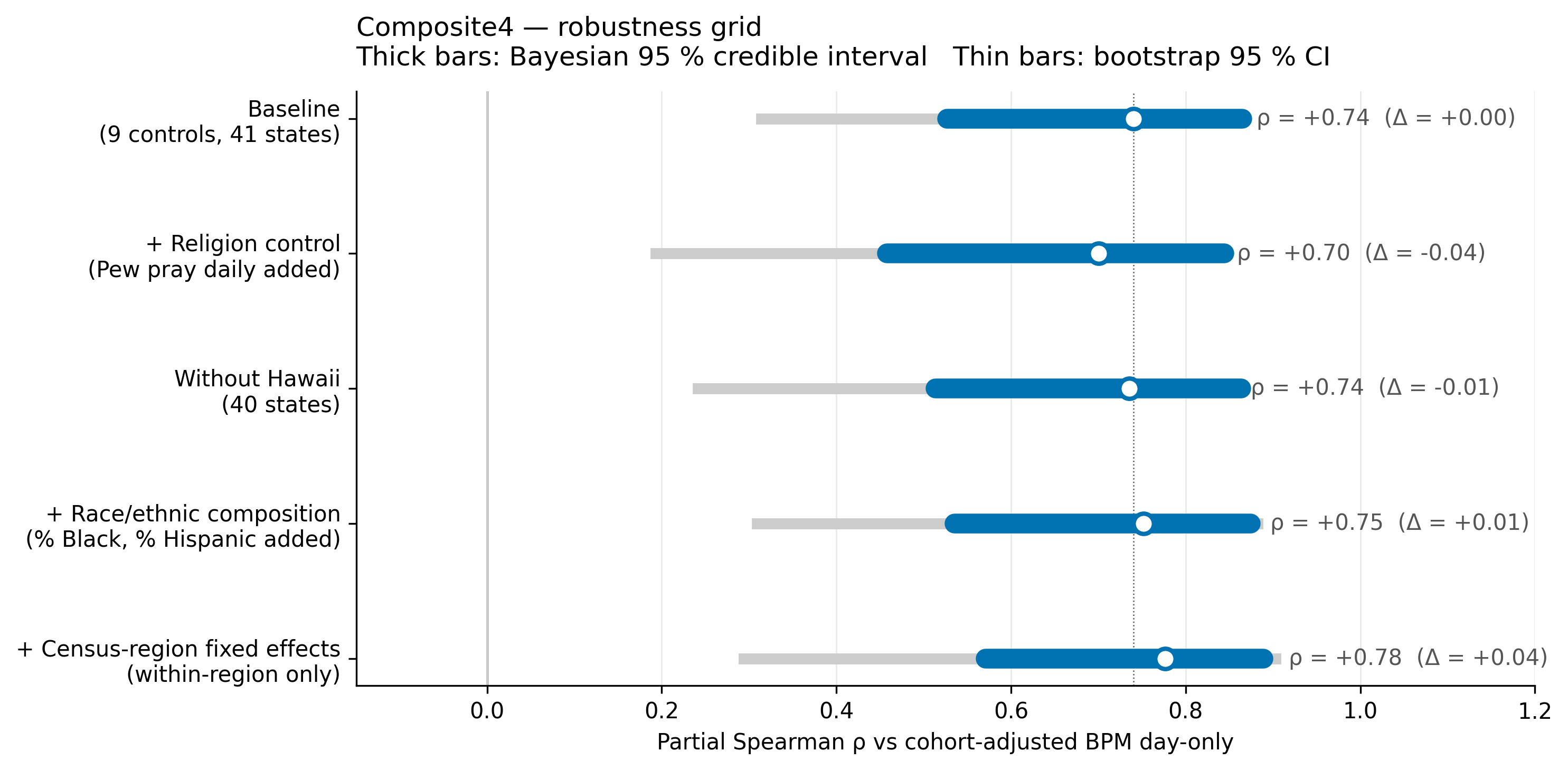}
\caption{Robustness grid. Partial Spearman $\rho$ (thick bars: Bayesian credible interval; thin bars: bootstrap CI) for the baseline specification ($+0.74$), with an added religiosity control ($+0.70$), excluding Hawaii ($+0.735$), with state racial and ethnic composition added as controls ($+0.752$), and with census-region fixed effects ($+0.776$). Every interval excludes zero.}
\label{fig5}
\end{figure}

Leave-one-state-out analysis confirmed that no single state drives the result: across all 41 leave-one-out fits the partial correlation remained in the range $+0.67$ to $+0.79$. The association was also insensitive to alternative specifications: adding a state-level religiosity control gave $+0.700$, excluding Hawaii (a geographic and demographic outlier) $+0.735$, and substituting Welltory-measured body-mass index for the CDC obesity-prevalence control $+0.713$ (Fig~\ref{fig5}).

The health and behaviour covariates can be read as confounders or as mediators, and the conclusion is the same either way. Read as mediators, a decomposition in bpm-per-SD units gives a total effect of $+0.66$ bpm/SD, a direct effect of $+0.53$ bpm/SD ($[+0.22, +0.94]$), and an indirect path through chronic disease of $+0.13$ ($[-0.20, +0.38]$) --- a mediated share of 17--20\% whose interval spans zero, so roughly 80\% of the association runs outside the health block. Read as confounders, adjusting for them raises the rank correlation from $+0.64$ to $+0.69$. These are different estimands on different scales --- the decomposition is a linear effect in bpm/SD conditioning on the non-mediator covariates, the rank partial a residual correlation conditioning on all nine --- which is why one reports a small mediated share and the other a slight strengthening; both say the gradient is not absorbed by chronic disease. Even adjusting for the single most hardship-collinear control in isolation leaves $+0.485$.

\subsection*{4.2\quad Geography and population composition}

Two structural features of a 41-state cross-section could produce a hardship--RHR gradient without any physiological relationship: spatial dependence (both variables cluster in the South, so the effective sample is smaller than 41 and the states are not independent draws) and the racial and ethnic composition of states, which is widely assumed to be entangled with area-level hardship in the United States. We tested both directly.

\textbf{Spatial dependence.} The raw variables are, as expected, strongly spatially clustered: Moran's $I$ on state mean resting heart rate is $+0.15$ under inverse-distance weights and $+0.35$ under 4-nearest-neighbour weights, and on the hardship composite $+0.13$ and $+0.48$ (all $p \leq 0.002$, 999 permutations, row-standardised weights over state centroids). The model residuals are not. Moran's $I$ on the residual of resting heart rate after the nine covariates is $-0.05$ (inverse-distance) and $-0.09$ (4-nearest-neighbour), and on the residual after the covariates \emph{and} the composite $-0.01$ and $+0.00$ --- in every case indistinguishable from the permutation expectation ($p \geq 0.45$). The covariate set absorbs the spatial structure; what the partial correlation is fitted on carries no detectable spatial dependence, though with 41 states this test has limited power against weak structure.

Consistent with that, the gradient is a within-region relationship rather than a South-versus-elsewhere contrast. Adding census-region fixed effects --- which discard all between-region variation --- \emph{raises} the partial correlation to $+0.776$ (bootstrap $[+0.29, +0.90]$). Dropping each region in turn leaves it at $+0.72$ without the South, $+0.64$ without the Midwest, $+0.66$ without the Northeast, and $+0.62$ without the West.

\textbf{Population composition.} At the state level, racial and ethnic composition is not the confounder it is often assumed to be here: the share of a state's population that is Black is essentially uncorrelated with the hardship composite ($\rho = -0.08$, $p = 0.62$), and the Hispanic share likewise ($\rho = +0.10$, $p = 0.54$). The White share is the one exception, correlating negatively with hardship ($\rho = -0.34$, $p = 0.028$) --- and adding it as a control raises the estimate rather than lowering it (below). The Black population share does correlate with cohort-adjusted resting heart rate, but negatively ($\rho = -0.36$, $p = 0.02$) --- the opposite sign from the one a compositional explanation of the gradient would require. Adding both shares to the control set moves the partial correlation from $+0.740$ to $+0.752$ (bootstrap $[+0.31, +0.88]$); adding the White share as well gives $+0.763$; composition together with region fixed effects gives $+0.791$. The last two specifications spend 12 and 14 degrees of freedom on 41 states, so we read them as directional confirmation rather than as sharper estimates. Composition is measured at the state level from ACS 2024, the same vintage as the median-age and density controls; the panel itself carries no individual race or ethnicity, so this addresses ecological confounding by composition, not individual-level effects, and it does not speak to the separate measurement-equity question raised by optical sensing (\secref{6}).

\subsection*{4.3\quad Device-mix sensitivity}

Because measurement hardware differs across users, a spurious association could in principle arise if device type both varied with state hardship and biased the physiological estimate. Apple Watch dominates the panel on every denominator: it supplies $\approx$98\% of all recordings, the median contributing user records essentially only on it, and $\sim$2\% of users record solely by smartphone camera. Averaged per user and then to the state, its share ranges from 0.85 to 0.96 (mean 0.92, SD 0.02) --- lower than the recording-weighted 98\% because the heaviest recorders are the most Apple-Watch-exclusive. That spread is enough for a sensitivity check, though not for a clean negative control, since the share is not independent of the exposure.

Device mix is not exogenous to the exposure: Apple-Watch share correlates with the hardship composite ($\rho = +0.36$, $p = 0.02$, 41 states), higher-hardship states skewing slightly more Apple-Watch within this opt-in panel. What closes the confounding path is the other leg --- share is only weakly related to the resting-heart-rate outcome ($\rho = +0.11$), and to the recovery metric Drain ($\rho = +0.10$), both inside the $|\rho| < 0.20$ gate. A path that barely reaches the outcome cannot carry the gradient, and adjusting for it directly confirms as much: adding Apple-Watch share as a tenth control leaves the partial correlation at $+0.726$ (from $+0.740$). The arousal-peak metric (Intensity) is the exception: it returned a non-negligible association with device share ($\rho = +0.31$), and for that reason Intensity is excluded from any state-level ranking or cross-sectional claim in this paper, while resting heart rate --- weakly related to device and unmoved by adjusting for it --- carries the headline result (Table~\ref{table3}).

\begin{table}[ht]
\centering\footnotesize
\caption{Reliability, device-mix, and multiplicity diagnostics. Device-mix rows correlate the state-level user-mean Apple-Watch (AW) share with each quantity (Spearman).}
\label{table3}
\begin{tabular}{L{4.6cm} L{4.0cm} L{6.4cm}}
\toprule
\textbf{Diagnostic} & \textbf{Value} & \textbf{Reading} \\
\midrule
\grouphead{3}{Reliability}
User-level split-half, day grain & Spearman--Brown 0.993 & Individual resting-HR measure is highly reliable \\
User-level split-half, month grain & 0.985 & Corroborates \\
State-ranking split-half (41 states) & 0.60 & Below 0.70 $\rightarrow$ no ordinal state-ranking claims \\
Year-over-year rank stability & 0.74 to 0.80 & Broad configuration reproduces, but shares users across years, so it is not an independent reliability estimate (\secref{4.4}) \\
\grouphead{3}{Device mix}
AW share, cross-state spread & 0.85--0.96 (mean 0.92, SD 0.02) & The device variable has real spread to test against \\
AW share $\times$ Composite4 & $\rho = +0.36$ ($p = 0.02$) & Tracks the exposure --- not exogenous \\
AW share $\times$ BPM day-only & $\rho = +0.11$ & Passes ($|\rho| < 0.20$) $\rightarrow$ gradient is not a device artefact \\
AW share $\times$ Drain & $\rho = +0.10$ & Passes $\rightarrow$ the device-clean discriminant metric \\
AW share $\times$ Intensity & $\rho = +0.31$ & Fails $\rightarrow$ Intensity excluded from state-level claims \\
Headline with AW share as 10th control & $+0.726$ (from $+0.740$) & Device is not the driver: the outcome leg closes the path \\
\grouphead{3}{Multiplicity}
Single-indicator screen & 1 of 112 survives strict Bonferroni & An unrelated correlation; the headline rests on holdout and nested CV, not Bonferroni \\
\bottomrule
\end{tabular}
\end{table}

\subsection*{4.4\quad Reliability at the user and state grains}

Reliability must be assessed at the grain of the claim. At the \textbf{user level}, the resting-heart-rate measure is highly reliable: a split-half analysis dividing each user's days into odd and even calendar days and correlating the resulting per-user means gave a Spearman--Brown reliability of 0.993 across the 17,379 users with at least 10 day-only readings in each half (for comparison, the recovery metric Drain reaches 0.887 at this same per-user, per-day grain --- the 0.887 is Drain's, not resting heart rate's); an alternative odd-versus-even-month split gave 0.985. The individual-level resting-heart-rate signal is thus among the most reliable the panel produces.

At the \textbf{state-ranking grain}, reliability is lower and bounds our claims accordingly. Splitting users into halves and correlating the resulting 41 state means gave a Spearman--Brown reliability of 0.60, below a conventional 0.70 threshold. This is why we make no ordinal state-ranking claims: individual state ranks are sample-limited at this scale. The gradient is instead asserted against the national mean with confidence intervals.

The state configuration also reproduces from one year to the next, at a rank correlation of 0.74 to 0.80 between 2024 and 2025. That figure exceeds the 0.60 split-half reliability, which at first looks impossible --- a retest correlation cannot exceed the reliability of what is being correlated. The two are not comparable: the split-half estimate divides the users of a single year into independent halves, whereas largely the same users contribute in both years, so any user-specific measurement error is shared between them and inflates the year-over-year figure. We therefore read 0.74--0.80 as evidence that the broad map is not a one-year accident, not as an independent estimate of state-mean reliability; 0.60 remains the operative bound on state-level precision (reliability and device-mix diagnostics are collected in Table~\ref{table3}).

\subsection*{4.5\quad Wearing-intensity sensitivity}

A prior version of this check defined heavy wearers by an absolute day-count threshold, which produced an uneven heavy/light composition across states and was withdrawn for that reason. In the corrected check, heavy wearers are defined \textbf{within each state} (above the state's own median day count), balancing the heavy/light split to roughly 50/50 in every state and removing the cross-state composition shift. On the within-state heavy-wear subset the partial correlation was $+0.475$ --- the same sign as the headline, attenuated in magnitude. Because the heavy subset contains roughly half the users per state, part of this attenuation reflects lower state-mean reliability at the smaller sample (\secref{4.4}) rather than genuine signal loss; the check's purpose is direction and stability under a balanced definition, both of which hold.

\subsection*{4.6\quad Multiple-comparisons context}

The four-component composite was selected from a larger space of candidate hardship and physiology combinations, and we report the association in that context rather than in isolation. In a broad screen of 112 single-indicator state-level associations, one survived a strict Bonferroni correction for the full set: a correlation between resting heart rate and a state religiosity measure (Pew daily-prayer share; marginal $\rho = +0.577$, $p = 9 \times 10^{-6}$, 51 jurisdictions; partial $+0.459$), unrelated to hardship and, added as a covariate, leaving the hardship association at $+0.700$ (\secref{4.1}). Most candidate signals were null.

The 112-screen is not, however, the operative multiplicity. Composite4 is the endpoint of a separate, iterative search over nine candidate composites drawn from a broader pool of indicators --- a combinatorially larger space than 112 single-indicator tests. The $-0.055$ nested-CV optimism (\secref{3.2}) quantifies component selection \emph{within} Composite4's four-indicator space, not that wider search. The operative multiple-comparisons scope is therefore the composite search itself, which is why the finding rests on internal split-sample validation, the robustness portfolio, and the specificity checks rather than on any significance threshold --- and why independent replication is the natural next step rather than a formality.

\section*{5\quad Discussion}

Passively collected consumer-wearable physiology tracked a known social gradient at population scale without administering a survey to the users. Across 41 US states, cohort-adjusted resting heart rate tracked a four-component measure of material hardship at a partial rank correlation of $+0.74$. The gradient held when the states were resampled, when the panel was re-weighted to external demographic margins, when state racial and ethnic composition was added to the controls, and when all between-region variation was removed; it was specific to resting heart rate rather than to any signal the panel produces, and was not matched by median income under the same adjustment. To our knowledge this is the first positive, cross-sectional, area-level demonstration of the resting-heart-rate--material-hardship relationship, and the first established from consumer-wearable data. The contribution is as much methodological as substantive: a passively sensed physiological signal, aggregated to the level of a state, carries information that tracks state-level material conditions.

\subsection*{5.1\quad Relation to established epidemiology}

The direction of the gradient is consistent with decades of cardiovascular epidemiology; its magnitude is best judged at the individual level rather than from the ecological correlation. At the individual level, resting heart rate rises with socioeconomic disadvantage: in the RECORD cohort it increased by roughly $+0.9$, $+1.8$, and $+3.6$ bpm across increasing categories of combined individual and neighbourhood disadvantage \cite{Chaix2011}. Our own user-level contrast --- $+1.33$ bpm between users in the five highest- and five lowest-hardship states (\secref{3.5}) --- is of the same order, whereas the ecological $+0.74$ correlation is, as expected, stronger than the individual-level association would be, and is not on the same scale as these bpm gradients. Higher resting heart rate is in turn a graded predictor of cardiovascular and all-cause mortality \cite{Aune2017,Cooney2010,Woodward2014}, and area-level socioeconomic deprivation predicts cardiovascular disease across a wide range of endpoints \cite{Pujades2014}. Our result connects these established individual-level and area-level findings through a new measurement channel: rather than clinical examination or survey linkage, the physiological signal is sampled passively from consumer devices, and rather than a disease endpoint, the social exposure is a direct composite of material hardship. That the expected gradient re-emerges through this channel is itself the finding.

\subsection*{5.2\quad A plausible pathway}

A well-developed physiological framework makes the association interpretable without invoking causation from these data. Chronic exposure to material hardship --- uninsurance, food insecurity, and utility and housing precarity --- is a source of sustained stress, and sustained stress maintains elevated sympathetic and reduced parasympathetic activity, which raises resting heart rate. The cumulative physiological cost of such prolonged activation is captured by the concept of allostatic load \cite{McEwen1998}, and lower socioeconomic status has been linked specifically to delayed cardiovascular recovery and heightened allostatic burden as a route to cardiovascular risk \cite{Steptoe2002}; the broader logic of physiological systems adjusting set-points in anticipation of demand --- allostasis --- provides the regulatory account \cite{Sterling2012}. We advance this as a plausible, previously documented pathway consistent with our cross-sectional, ecological association --- not as evidence of causation, which our design cannot establish.

One structural alternative is worth stating precisely, because it is usually assumed rather than tested: that an area-level hardship gradient is really the racial and ethnic composition of states in disguise. In these data it is not. The Black and Hispanic population shares are uncorrelated with the hardship composite ($\rho = -0.08$ and $+0.10$); the White share does correlate ($-0.34$), but adjusting for it, or for all three, raises the estimate rather than lowering it, and the composition--RHR association runs opposite in sign to what the compositional explanation requires (\secref{4.2}). Adding the state-level composition variables we tested therefore did not attenuate the association; it says nothing about individual-level pathways, which the panel cannot address, and nothing about the separate measurement-equity question of how optical sensing performs across skin tones (\secref{6}). Material hardship at the state level remains correlated with other structural factors we have not measured, so the gradient should not be read as isolating one specific mechanism.

\subsection*{5.3\quad Reconciliation with a prior opposite-direction result}

The one prior study we are aware of that related area-level deprivation to heart rate reported an association in the \emph{opposite} direction --- decreasing heart rate with greater neighbourhood disadvantage --- in the Baltimore Study of Black Aging \cite{Allan2024}. Several design differences reconcile this with our result rather than placing it in tension. First, that finding is longitudinal, describing \emph{change} in heart rate across two waves roughly three years apart, whereas ours is a cross-sectional estimate of the \emph{level} of resting heart rate; the two need not share a sign. Second, it was obtained in a demographically narrow cohort --- older adults in a single population --- that differs from our general panel, which limits direct comparability; we do not attribute the sign difference to any single population characteristic, and note that age-related compression of heart-rate range and medication use in an older cohort can attenuate or reverse chronotropic gradients. Third, it derives from a single-city sample of 317 participants, versus our 41-state cross-section. Rather than contradicting the present result, the comparison suggests the direction of the deprivation--heart-rate relationship is sensitive to study design and population, and identifies our estimate as the first cross-sectional area-level estimate of resting-heart-rate \emph{level} across a general consumer-wearable panel.

\subsection*{5.4\quad Implications for population physiological sensing}

The practical implication is that opt-in consumer-wearable panels, though non-representative, can surface population-level structure that agrees with external indicators --- here, the geography of material hardship. Passive physiological data are repeated, timely, and inexpensive relative to survey instruments, so they could complement conventional social and health surveillance where those instruments are slow or coarse. They are a complement rather than a replacement: what makes such a panel usable is the external-consistency and specificity checking applied here, and the non-representativeness stated openly alongside the estimate.

\subsection*{5.5\quad Limitations in brief and outlook}

The expected social gradient in resting heart rate re-emerges from passively sensed data, through a channel entirely different from the clinical and survey instruments that first established it. The natural next steps are replication in other consumer-wearable panels, temporal analyses relating within-state changes in hardship to changes in resting heart rate, and linkage designs that could separate the contributions of the individual and the area. What bounds the present estimate is its design, not its stability: it is ecological, cross-sectional, drawn from an opt-in panel, and built on a composite selected within a multiple-comparisons search, so it licenses no causal or individual-level inference (full limitations in \secref{6}).

\section*{6\quad Limitations}

Our study has several limitations, which bound the claims above and which we state directly.

\textbf{The panel is opt-in and not population-representative.} Users self-select into the application, and their geographic and demographic distribution does not match the US adult population; in particular, the panel under-samples the highest-hardship states. The direct-standardisation analyses (\secref{3.3}) refine rather than remove this limitation. The direction of the selection effect is unknown: depending on how app uptake, device ownership, health interest, and measurement frequency interact, it could attenuate, inflate, or leave the gradient unchanged, so we do not claim the estimate is conservative.

\textbf{The outcome is an internally derived resting-heart-rate estimate, not a clinical resting measurement.} Our measure is the per-user median of short, opportunistically scheduled daytime photoplethysmography readings (\secref{2.2}); no single reading is a resting-state measurement, and the resting-level interpretation rests on the median over many readings approximating a typical waking level. We have not validated this estimate against a conventional resting heart rate --- a clinician-measured seated value or a device-reported nocturnal-sleep minimum --- so its correspondence to the ``resting heart rate'' of the epidemiological literature (\secref{1}, \secref{5.1}) is an assumption rather than a demonstrated equivalence. A direct test against a nocturnal-sleep resting heart rate is not possible within this dataset, which is waking-only: the application's night readings are pre-sleep quiet-waking, and its sleep signal is a separate pipeline not included here. Such a validation is a natural next step that would require that additional data; the RECORD comparison (\secref{5.1}) and the mortality priors (\secref{1}) should be read with this estimate's construct status in mind.

\textbf{The estimate is ecological and cross-sectional, and supports neither individual-level nor causal claims.} The $+0.74$ correlation is a state-level (area-level) relationship describing how states differ, not how individuals differ. Aggregation removes within-area variation and can make an area-level association either stronger or weaker than its individual-level counterpart; in either case it cannot be translated into an individual-level estimate, and inferring individual behaviour from area-level relationships is a well-known fallacy \cite{Robinson1950}. The within-state, between-user standard deviation in resting heart rate ($\approx$ 9 bpm) far exceeds the between-state spread; the $+1.33$ bpm user-level contrast (\secref{3.5}) is the only individual-level quantity we report, and it should not be read as an individual-level effect size or as a statement about any person's cardiovascular risk. The design is equally silent on causation: we report an association consistent with an established physiological pathway (\secref{5.2}), but nothing in the data identifies a causal direction or rules out unmeasured common causes. The finding is descriptive, not explanatory.

\textbf{The composite was selected post hoc within a multiple-comparisons search.} The four-component composite was not the pre-specified primary metric (\secref{2}). In a screen of 112 single-indicator associations only one survived a strict Bonferroni correction --- an unrelated correlation, not the composite (\secref{4.6}) --- so we do not rest the finding on nominal significance. We treat it as hypothesis-generating and quantify the selection optimism by nested cross-validation (\secref{3.2}), and independent replication remains necessary.

\textbf{State ranks are sample-limited, and state means are treated as equally precise.} State-ranking reliability is 0.60, below a conventional threshold (\secref{4.4}), so we make no ordinal state-ranking claims and do not interpret the position of any individual state; the gradient is asserted against the national mean with confidence intervals. The correlation also weights every state equally, though the number of contributing users ranges from 75 to 1,668, so state means differ in precision. We propagate that measurement error into the uncertainty through the two-stage bootstrap (\secref{2.6}, \secref{3.1}), which widens the interval to $[+0.06, +0.79]$, but we do not correct the point estimate for it: a precision-weighted or errors-in-variables estimator remains the natural refinement, and it would be expected to move the estimate upward, in the opposite direction from the widened interval.

\textbf{Individual race and ethnicity are absent, so composition is addressed only ecologically.} The panel carries state, sex, and age but not race or ethnicity, and our cohort adjustment and standardisation are correspondingly limited to sex and age. We can and do adjust for racial and ethnic composition at the \emph{state} level, where it is publicly measured, and the gradient is unchanged by it (\secref{4.2}). What remains untestable here is the individual level: whether the hardship--RHR relationship differs across racial and ethnic groups within states, and whether individual-level composition effects offset or amplify the area-level pattern. Those questions require individual-level data the panel does not contain.

\textbf{Measurement is single-sensor, and optical-sensor equity is unaddressed.} Readings come overwhelmingly from Apple Watch photoplethysmography ($\approx$98\% of recordings), with a small smartphone-camera minority ($\sim$2\% of users record only by camera), so the panel is closer to a single instrument than to a heterogeneous mix. Device heterogeneity is therefore a minor source of measurement noise rather than a dominant one; our headline metric passes the device-mix sensitivity check and is essentially unchanged when Apple-Watch share is added as a control (\secref{4.3}), while one ancillary metric (Intensity) fails that check and is excluded from cross-sectional claims for that reason. Being almost entirely one optical-PPG device limits device heterogeneity but raises a separate concern it does not neutralise: optical photoplethysmography signal quality can depend on skin pigmentation as well as on motion, perfusion, and device conditions, and a single optical sensor is not skin-tone-independent. The panel carries no race, ethnicity, or skin-tone information, so we cannot test this. Racial and ethnic composition varies across states whether or not it tracks hardship (\secref{4.2}), so any skin-tone dependence in quality-control pass rates would make the set of users retained per state vary systematically, and could distort the geographic pattern even in the absence of a true physiological difference. A skin-tone-stratified analysis of signal quality and dropout is impossible without skin-tone data; this remains an unquantified limitation and a necessary next test. Accordingly, no racial or ethnic physiological interpretation of the state pattern is possible or intended, and the state-level estimates must not be used to classify individuals or groups or to inform eligibility decisions.

\textbf{Much of the external data is model-based, shares one survey source, and is in part single-vintage.} Two of the four hardship components and all six health-and-behaviour controls are CDC PLACES small-area model-based estimates derived from BRFSS --- modelled rather than directly measured, and sharing a common estimation pipeline. The utility-shutoff and housing-insecurity components come from the PLACES 2022 BRFSS Social Determinants and Health Equity module, fielded only in 2022 (\secref{2.4}); unlike the outcome, whose year-over-year rank stability we report (\secref{4.4}), we cannot check whether these two indicators' state rankings are stable across years or were distorted by 2022-specific conditions, and the two correlate at Spearman $\approx 0.81$, making them two facets of one module rather than independent confirmations. Neither issue can manufacture the headline: the outcome (resting heart rate) comes from an entirely independent consumer-wearable source, so shared-method correlation among the BRFSS-derived variables cannot create the hardship--resting-heart-rate association. Consistent with this, the association is already positive without any BRFSS controls (the marginal correlation, $+0.64$) and without the two BRFSS-derived components (Composite2, $+0.57$); no single state drives it (leave-one-state-out $+0.67$ to $+0.79$); and each module component individually lifts the composite (adding either alone raises the two-component partial from $+0.634$ to $+0.719$ and $+0.711$ respectively).

\section*{7\quad Data and code availability}

Aggregate data and analysis code sufficient to reproduce the state-level statistical analyses reported in this paper from the released aggregate tables are openly available at Zenodo (DOI: \href{https://doi.org/10.5281/zenodo.21391207}{10.5281/zenodo.21391207}), under CC-BY-4.0 for data and the MIT licence for code. The deposit separates the wearable-derived state aggregates --- cohort-adjusted resting heart rate and two ancillary metrics per US state, each averaged over at least 75 users --- from the public US-government indicators used as composite components and controls (US Census / American Community Survey, USDA Economic Research Service, CDC PLACES, and related federal sources), and includes a script that joins them and regenerates the reported statistics. Per-state distributional summaries (mean, median, standard deviation, and bootstrap confidence interval) are provided; states enter only with at least 75 contributing users, and any finer cell in the deposit (for example, a state $\times$ time-period breakdown) with fewer than 15 users is suppressed. No per-user table is released: because the underlying measurements are individual physiological data from an opt-in, non-representative panel, we release state-level aggregates rather than de-identified individual rows.

Everything estimated at the state level reproduces from the deposit: the headline partial correlation and both intervals, the covariate decomposition, leave-one-state-out, the spatial and composition checks, the holdout and nested cross-validation, the demographic re-weighting, and the specificity comparisons. Three results are computed from per-user data and therefore cannot be regenerated from the public release: the user-level split-half reliabilities (\secref{4.4}), the within-state heavy-wear subset (\secref{4.5}), and the individual-level anchor (\secref{3.5}). Their inputs are described in the sections cited and their outputs are reported in full above.

\section*{8\quad Ethics statement}

This work is a secondary analysis of pre-existing physiological measurements collected by Welltory Inc.\ in the ordinary course of operating its consumer health application. All contributing users are adults; the application is not offered to minors, and no under-18 records enter the panel. The Terms of Service and Privacy Policy in force throughout the 2024--2025 measurement window informed users that Welltory may process measurements for research, analytics, service-improvement, and app-functionality purposes, and may use or share aggregated and/or anonymized, statistically re-worked data that cannot reasonably be used to identify an individual, including for research and open-science publications. No intervention, no new data collection, and no contact with participants were undertaken for this study; individual measurements were accessed internally by one author (M.L.), in the researcher role under contract to Welltory described in Competing interests, solely to generate the state-level aggregate estimates and analyses reported here. No independent ethics body reviewed this study: no institutional review board, research ethics committee, or equivalent external reviewer issued a determination, and the judgement that the analysis fell within the existing Terms of Service and Privacy Policy was made internally. We state this plainly rather than imply a review that did not occur, and we note that an internally made determination is weaker than an independent one: the fact that data are pre-existing and company-held does not by itself establish that independent review was unnecessary. An independent determination is the appropriate next step, and would be required before any extension of this work. Only state-level aggregates are publicly released as described in \secref{7}; the release contains no raw physiological measurements, per-user records, persistent identifiers, individual-level datasets, or precise geolocation data, and is designed so that it cannot reasonably be used to identify any individual user.

\section*{Author contributions}

M.L.\ designed the analysis, built the data pipeline, ran all statistical analyses, produced the figures, and wrote the manuscript. V.A.\ contributed to the analysis design and reviewed the statistical approach. J.S.\ contributed to framing the research question and reviewed the manuscript. All authors approved the submitted version.

\section*{Funding}

This work received no external funding. It was carried out with internal resources of Welltory Inc., which employs or contracts the authors (see Competing interests). No funder outside the authors' own institution had any role in the study design, analysis, decision to publish, or preparation of the manuscript.

\section*{Competing interests}

All authors are affiliated with Welltory Inc., which developed the application and owns the aggregated physiological data analysed here (M.L.\ as an independent contractor researcher; V.A.\ as Head of AI and J.S.\ as founder). These affiliations are the study's principal competing interest and are declared accordingly. Readers should weigh them against the record rather than against any assurance from us: the primary metric and the train--holdout split were specified before the construct refinement, the four-component composite was developed post hoc in an explicitly exploratory analysis (\secref{2}, \secref{4.6}), the pre-specified index's weaker result is reported alongside it, and the state-level aggregates and analysis code are released (\secref{7}) so the analyses can be checked independently. Welltory has authorized public release of the manuscript, the state-level aggregate dataset, and the analysis code under the licenses stated in the Data and code availability section.

\section*{Acknowledgements and disclaimers}

This study uses publicly available federal data from the US Census Bureau / American Community Survey, USDA Economic Research Service, CDC PLACES, and related federal sources. Use of these data does not imply endorsement by CDC, USDA, the US Census Bureau, HHS, or the United States Government.

Apple Watch is a trademark of Apple Inc., registered in the U.S.\ and other countries and regions. Apple Inc.\ was not involved in the study and does not endorse the findings.

The wearable-derived metric and state-level aggregate estimates are reported for research purposes only and are not intended or validated for diagnosis, treatment, clinical decision-making, individual cardiovascular-risk assessment, or eligibility decisions, including decisions related to employment, insurance, lending, housing, healthcare access, or public benefits.

\bibliographystyle{unsrt}
\bibliography{references}

\appendix
\section*{Supporting information}

\begin{table}[ht]
\centering\footnotesize
\caption*{\textbf{S1 Table.} Composite search: candidate hardship composites. Partial $\rho$ is adjusted for the covariate set current in each run; the EPI 51-state partials are as reported in their source runs and are not strictly comparable, covariate-for-covariate, to the finalised 41-state figures.}
\setlength{\tabcolsep}{4pt}
\begin{tabular}{L{2.3cm} L{5.1cm} c c L{3.8cm}}
\toprule
\textbf{Composite} & \textbf{Components} & \textbf{n} & \textbf{Partial $\rho$} & \textbf{Status} \\
\midrule
EPI v1 & unemployment $+$ ACS rent burden 30\%$+$ $+$ uninsured $+$ food insecurity & 51 & --- (housing input error) & Housing input affected by a data error, corrected in later runs \\
EPI v2 & unemployment $+$ uninsured $+$ housing insecurity $+$ food insecurity & 51 & $+0.421$ & Includes unemployment, a labour-market condition rather than a direct hardship experience \\
EPI v3 & unemployment $+$ uninsured $+$ ACS rent burden 30\%$+$ $+$ food insecurity & 51 & $+0.292$ & Pre-specified specification; rent burden also captures high-cost housing markets, not deprivation \\
EPI v4 & unemployment $+$ uninsured $+$ ACS rent burden 50\%$+$ $+$ food insecurity & 51 & $+0.301$ & Severe-burden threshold does not resolve the high-cost-market confound \\
Composite2 & uninsured $+$ food insecurity & 51 & $+0.573$ & Retained (national-coverage supplement) \\
Composite2 (matched) & uninsured $+$ food insecurity & 41 & $+0.634$ & Augmentation base for the 41-state construct refinement \\
Composite3h & Composite2 $+$ housing insecurity & 41 & $+0.719$ & Intermediate --- housing augment \\
Composite3s & Composite2 $+$ utility shutoff & 41 & $+0.711$ & Intermediate --- utility augment \\
\textbf{Composite4} & uninsured $+$ food insecurity $+$ utility shutoff $+$ housing insecurity & 41 & $+0.740$ & \textbf{Exploratory headline} --- direct experienced-hardship indicators only \\
\bottomrule
\end{tabular}
\end{table}

\begin{table}[ht]
\centering\footnotesize
\caption*{\textbf{S2 Table.} Selected single-indicator candidates, including the one Bonferroni survivor (religiosity) and the nine most informative retirements. Partial $\rho$ is against BPM day-only, adjusting for the covariate set current at screen time. These are illustrative of, not exhaustive of, the 112 single-indicator associations referred to in \secref{4.6}; the full screen is in the Zenodo deposit (\secref{7}).}
\begin{tabular}{L{3.8cm} L{2.8cm} L{2.9cm} L{5.3cm}}
\toprule
\textbf{Candidate} & \textbf{Source} & \textbf{Partial $\rho$ vs BPM} & \textbf{Outcome} \\
\midrule
Pew daily-prayer share & Pew RLS 2023--24 & $+0.46$ (marginal $+0.58$) & \textbf{Only indicator surviving strict Bonferroni}; unrelated to hardship, retained as a robustness covariate (\secref{4.1}) \\
BLS state unemployment 2025 & BLS LAUS & $+0.33$ & Labour-market condition rather than direct experienced hardship \\
ACS rent burden 30\%$+$ & US Census B25070 & $-0.14$ & Captures high-cost housing markets rather than deprivation \\
ACS rent burden 50\%$+$ (HUD severe) & US Census B25070 & $-0.03$ & Severe-burden threshold does not isolate deprivation \\
HPS rent late-or-missed & US Census HPS Cycle 09 & $-0.16$ & High-cost-market confound \\
HPS eviction likelihood & US Census HPS Cycle 09 & $+0.03$ & Not retained \\
HPS foreclosure likelihood & US Census HPS Cycle 09 & $+0.10$ & Not retained \\
CDC PLACES \texttt{foodinsecu} & CDC BRFSS SDOH module & $+0.29$ & Redundant with USDA food insecurity \\
Bankruptcy per 100k & US Courts F-5A FY2025 & $+0.05$ (BPM); $+0.31$ (Drain) & Not in the BPM composite --- exploratory Drain finding \\
Overdose per 100k & CDC NCHS & $-0.03$ & Not retained \\
MHA suicide-ideation rate & Mental Health America & $-0.25$ & Reflects help-seeking self-selection \\
\bottomrule
\end{tabular}
\end{table}

\end{document}